\documentclass[twocolumn,showpacs,preprintnumbers,amsmath,amssymb]{revtex4}

\usepackage{graphics}
\usepackage{epsfig}

\begin{document}
\def\am{{$a_{\mu}~$}}
\def\amp{{$a_{\mu ^+}~$}}
\def\amm{{$a_{\mu ^-}~$}}
\def\omegaa{{$\omega_a~$}}
\def\omegap{{$\omega_p~$}}
\def\ecm{\rm e \cdot cm}
\def\ni{\noindent}
\def\st{\scriptstyle}
\def\sst{\scriptscriptstyle}
\def\mco{\multicolumn}
\def\epp{\epsilon^{\prime}}
\def\vep{\varepsilon}
\def\ra{\rightarrow}
\def\um{\rm  \mu  m}
\def\us{\rm  \mu  s}
\def\al{\alpha}
\def\ab{\bar{\alpha}}
\def\bea{\begin{eqnarray}}
\def\eea{\end{eqnarray}}
\def\oma{\omega_a}
\def\omb{\omega_b}
\def\ma{magnetic}
\def\el{electric}
\def\f{field}
\def\fs{fields}

\title{A new method of measuring electric dipole moments in storage rings}
\author{
F.J.M.~Farley$^{7}$,
K.~Jungmann$^4$,
J.P.~Miller$^2$,
W.M.~Morse$^3$,
Y.F.~Orlov$^5$,
B.L.~Roberts$^2$,
Y.K.~Semertzidis$^3$,
A.~Silenko$^1$,
E.J.~Stephenson$^6$
}

\affiliation{
\mbox{$\,$}\\
\mbox{$\,^1$Belarusian State University, Belarus};
\mbox{$\,^2$Physics Department, Boston University, Boston, MA 02215};
\mbox{$\,^3$Brookhaven National Laboratory, Upton, NY 11973};
\mbox{$\,^4$Kernfysisch Versneller Instituut, Groningen};
\mbox{$\,^5$Newman Laboratory, Cornell University, Ithaca, NY 14853};
\mbox{$\,^6$IUCF, Indiana University, Bloomington, IN 47408};
\mbox{$\,^{7}$Department of Physics, Yale University, New Haven, CT 06520}.\\
}

\date{\today}

\begin{abstract}

A new highly sensitive method of looking for electric dipole moments of charged 
particles in storage rings is described.  
The major systematic errors inherent in the method  are addressed and ways to minimize them
are suggested.
It seems possible to measure the muon EDM to levels that  test speculative
theories beyond the standard model.

\end{abstract}

\pacs{13.40.Em, 12.60.Jv, 14.60.Ef, 29.20.Dh}

\maketitle

The existence of a permanent electric dipole moment (EDM) for an elementary 
particle would violate parity (P) and time reversal 
symmetry (T) \cite{tinv}. Therefore under the assumption of CPT invariance, a 
non-zero EDM would signal CP violation. In the standard model, the electron EDM is 
$< 10^{-38} \ecm$ \cite{smedm} with the muon EDM scaled up by the
mass ratio $m_\mu / m_e$, a factor of  $207$, 
 but some new theories predict much larger
values~\cite{pilaftsis,babu}. 
For example, ref. \cite{babu} predicts the muon EDM could be as
large as $5\times 10^{-23} \ecm$, while the electron EDM is predicted to be
$\sim 10^{-28} \ecm$, an order of magnitude below the present limit~\cite{commins02}.  The
current 95\%  confidence limit for the muon EDM is $10^{-18} \ecm$~\cite{cern2}.  
This paper discusses a new way of using a
magnetic storage ring to measure the
EDM of the muon,
which also can be applied to other charged particles.

To measure the EDM experimentally, the particle should be in an electric
field which exerts a torque on the dipole and induces an observable precession
of its spin. If the particle is charged this electric field inevitably
accelerates the particle; it will move to a region where the field is zero
or leave the scene. 
An example is the nucleus at the center of an atom in equilibrium;
the net force and therefore the net electric field
at the nucleus must average to zero according to Schiff's theorem~\cite{schiff}. 
Any applied external
electric field will be shielded from the nucleus by the electrons in the atom.
The overall effect is to suppress the EDM signal, making it more difficult
to measure. The suppression would be total but for
the many known exceptions to Schiff's theorem
when weak and strong forces, weak electron-nucleon forces, finite particle
sizes, and relativistic effects are included.
Suppression of the EDM signal by Schiff's theorem is completely avoided
in a magnetic storage ring~\cite{garled,Sandars_99} such as proposed here,
because the particle is not in equilibrium; there is a net centripetal
force, and this force is entirely supplied by a net electric field
as seen in the muon rest frame.

In particular, when a muon of velocity $\vec \beta=\vec v /c$
and relativistic mass factor $\gamma=(1-\beta^2)^{-{1\over 2}}$ is circulating
in a horizontal plane due to a vertical magnetic field $\vec B$,
it will according to a Lorentz transformation experience both an electric and
a magnetic field, $\vec E^*$ and $\vec B^*$, in  its own
rest frame.
The so-called motional electric field,
$\vec E^*=\gamma c \vec \beta \times \vec B$, can be much larger than any
practical applied electric field. Its action on the particle supplies the
radial centripetal force, Thomas spin precession, and
spin precession due
to any non-vanishing EDM. $\vec B^*$ produces precession due to the
muon magnetic moment.
The combined spin precession due to the Thomas precession and 
torque on the magnetic moment is vertically directed and is given~\cite{bmt,jackson},
in terms of the laboratory frame field, by
$\omega_a=a(e B/m)$, where $a=(g-2)/2$ is the
magnetic anomaly. This is referred to as the ``$(g-2)$'' precession, and if it
were acting by
itself it would
cause spin precession in the horizontal plane. $\omega_a / 2 \pi$ is referred to as the
``$(g-2)$'' frequency.
If there is an EDM of magnitude $d=\eta e \hbar  /4mc\approx \eta\times 4.7
\times 10^{-14} \, \rm e \cdot cm$, there will be
an additional precession angular frequency
\begin{equation}
\vec{\omega} _e \ =\   {\eta \over 2} {e \over m}\ \vec{\beta} \times \vec{B}\label{eq:A}
\end{equation}

\noindent
about the direction of $\vec E^*$, that is in the  radial direction
with respect to the orbit~\cite{garled}.
The vector combination of $\vec \omega_a$ (vertical) and $\vec \omega_e$
(radial) tilts the precession plane out of the horizontal plane, leading
to a vertical component of spin which oscillates at the frequency
$\omega=\sqrt{\omega_a^2+\omega_e^2}$ with an amplitude proportional
to the EDM. The decay electron direction is correlated with the spin
direction;
therefore the decay electrons acquire a small oscillating vertical
component of momentum. A search for this oscillation during the CERN $(g-2)$
experiment~\cite{cern1}  led to the current muon EDM limit~\cite{cern2}  
which corresponds to $\omega_e/\omega_a \leq 10^{-2}$.

This method has been used during the muon $(g-2)$ experiment at 
Brookhaven~\cite{bnl}, but  is limited by 
serious systematic effects.  First, because of the $(g-2)$ rotation, the EDM effectively acts for only one quarter of the $(g-2)$ period, thereby reducing the EDM signal.
 Also  the two extremes of 
the vertical oscillation occur when the muon spin is aligned radially 
inwards and outwards.  In these two extremes the decay electrons, 
whose up/down asymmetries are to be compared, follow rather different 
tracks through the magnetic field.  In one case the majority are 
emitted radially inwards and take a short path to the detectors; 
while in the other case they are emitted predominantly outwards and 
reach the detectors after a longer track with more opportunity to 
spread vertically or to be bent by stray radial magnetic fields.  The 
horizontal $(g-2)$ precession thus interferes with attempts to 
observe the vertical precession due to the EDM.

The new technique is to cancel the $(g-2)$ precession $\vec \omega_a$ so that the
radially directed $\vec \omega_e$ can operate by itself.
For a particle whose spin is initially
polarized along its momentum direction, the spin will rotate about the
radial direction, acquiring a vertical component,
so that the angle between the spin and the horizontal (orbit) plane increases
from zero linearly with time. Instead of a small vertical spin component
oscillating above and below the plane,
we now have a vertical spin component which can, by comparison, become
quite large with time, thereby greatly enhancing the EDM signal.
The cancellation can be achieved by applying a strong radial electric field
$\vec E$ to the orbit.
The expression for the angular velocity vector 
of the $(g-2)$ precession~\cite{bmt,jackson} in the presence of an $\vec E$
as well as a $\vec B$ field is, with 
$\vec \beta\cdot \vec E=\vec \beta\cdot \vec B=0$,
\begin{equation}
\vec{ \omega} _a=\frac{e}{m}\left[ a\vec{B}+(\frac{1}{\beta^2\gamma^2}-a)\ 
\vec{\beta} \times \vec{E}/c \right]       \label{eq:B} 
\end{equation}
valid for both fermions and bosons.
The latest muon $(g-2)$ experiments~\cite{cern1,bnl} run at the ``magic'' momentum of 3.1~GeV/$c$ with 
$\gamma \approx 29.3$ where the second term of equation~\ref{eq:B} vanishes 
 and $\vec{ \omega} _a=   a(e \vec{B} /m) $.  For the dedicated EDM experiment 
we are proposing to use muons with
momentum below the ``magic'' momentum value. 
If $1/(\beta^2 \gamma^2) \gg a$ and the electric field is adjusted to 
\begin{equation}
E\ =\ E_0\ \approx \  a Bc\beta \gamma^2       \label{eq:C} 
\end{equation}

\noindent $\omega _a$ can be reduced to zero (see also~\cite{nelson59}).   
The correct value can be set in the laboratory by monitoring the 
 cancellation of the 
$(g-2)$ precession with electron detectors on the inside of the ring. 
Then  $\vec{\omega}_e$ in equation (\ref{eq:A}) 
will have its full effect, moving the spin steadily 
out of the horizontal plane.
The vertical asymmetry can be observed with 
detectors, located  above and below the orbit, to measure the EDM without the 
systematic errors mentioned above.

To obtain the best accuracy it is desirable to use a high magnetic 
field and high energy muons which live longer.
But 
equation~(\ref{eq:C})  shows that this would require impractically large 
electric fields.  The parameters of a possible experiment are shown in Table I.

\bigskip
\begin{centerline}
{Table I.  Parameters of a possible muon EDM experiment.}
\begin{center}
\begin{tabular}{|c|c|c|c|c|c|} \hline

    E      & Aperture & B & p & $\gamma \tau$ & R \\
\hline
    2 MV/m   & 0.1~m & 0.25~T & 0.5 GeV/c & $11 \us$ & 7~m \\
\hline

\end{tabular}
\end{center}
\end{centerline}

\ni The uncertainty in $\eta$ is 
\begin{equation}
\sigma_\eta = {\sqrt{2} \over \gamma \tau ({e/m}) \beta B A P \sqrt{N}}, \label{eq:err}
\end{equation}

\ni where $A$ is the vertical asymmetry of the detected electrons for 100\%
 muon beam polarizarion  and $P$ is the actual muon beam polarization.  N is 
the total number of detected electrons, and $\tau$ is the muon lifetime at rest.

For example, to reach the sensitivity of  $10^{-24}\ecm$ in the EDM
corresponding to $\omega_e /\omega_a\ =\ 10^{-8}$,  the 
 vertical spin angle to be measured after $3$ dilated lifetimes ($33 \us$) 
would be $\sim 50$~nR, generating a counting asymmetry of $10^{-8}$ 
and requiring about $4 \times 10^{16}$ registered events, assuming $A=0.3$
and $P =0.5$, i.e. $N P^2 = 10^{16}$.  
Reaching  $10^{-24}\ecm$ would require 
a high intensity muon source plus a storage ring of large acceptance.  The muon EDM
collaboration has
submitted a letter of intent to J-PARC~\cite{muon_edm} where the requisite muon
beam line has been proposed.

In practice detectors, called $(g-2)$ detectors, would be set up to monitor the $(g-2)$ precession 
(horizontal spin motion).  Other detectors, above and below the orbit
 called EDM detectors, would be set up to 
measure the vertical spin motion.  With the electric field set to 
 some value below $E_0$ one can observe the $(g-2)$ precession and determine its
precession plane. 
As $E$ approaches $E_0$ the 
$(g-2)$ frequency $\omega_a / 2 \pi$ will gradually decrease.  
At the same time the EDM signal should 
have the same period as  $(g-2)$ but its amplitude should grow.  When 
the $(g-2)$ motion is cancelled the EDM signal should grow 
linearly with time.  In principle $\omega_e$ could make the spin turn several times 
in the vertical plane. But if the EDM is small, as expected, one will only observe
the beginning of the first oscillation, that is a slow linear rise.

Any storage ring must have horizontal and vertical focusing (quadrupoles or magnetic 
gradients) to keep the particles in orbit and the particles will in general
make betatron oscillations about the equilibrium orbit which will not necessarily
be exactly flat or in one plane.  
A number of imperfections in the magnetic or electric fields would 
make the spin move  out of the plane of the orbit 
 even though the EDM is zero, giving rise to a 
false EDM signal.  In the discussion ``the horizontal plane'' means the plane of the orbit,
while ``vertical'' and subscripts ``V'' refer to components normal to the plane of the orbit.
The following imperfections have been considered:
\begin{enumerate}

\item Vertical corrugations of the orbit due to a radial magnetic 
field $B_r$.

\item The plane of the radial electric field 
does not coincide with the plane defined by the magnetic field (called the 
``magnetic plane''), that is $E_{\rm v} \neq \ 0$ although the electric field is 
perfectly in a single plane so $\langle E_{\rm v} \rangle \ = \ 0$ with the brackets 
$\langle \ \rangle $ 
indicating the average over the orbit.

\item The electric field is not in one plane, $\langle E_{\rm v}\rangle \ \neq \ 0$.

\item Local orbit distortions near the detectors simulate detector rotation 
around the beam direction, so small residual $(g-2)$ precession (``horizontal'') 
has a component in the ``vertical'' direction looking like a false EDM.

\item Change of up detector response relative to down detector response
 during the muon storage time.

\item Azimuthal components $B_\theta$ of the magnetic field parallel to 
the momentum vector $\vec{p}$.  Although $\langle B_\theta\rangle \ =\ 0$ if there is 
no electric current through the orbit, higher harmonics of the azimuthal
B-field could be significant.

\end{enumerate}

We will discuss these effects in turn.
In any ring structure with \ma\ and/or \el\ \fs\,, for each particle 
momentum there exists a closed orbit; the particle repeats this track 
perfectly from turn to turn.  Other particles, starting at different 
transverse positions or transverse angles, oscillate about the closed 
orbit.  To define the plane of the closed orbit, split it up into many 
small equal sections, each with its local angular velocity vector 
$\vec{\omega}$ and find the average value $\langle \vec{\omega}\rangle $.  The orbit 
plane is defined as the plane perpendicular to $\langle \vec{\omega}\rangle $; on 
average the momentum vector rotates in this plane but may oscillate 
above and below it.  
With only a magnetic field, the orbit plane will therefore be defined by the 
average direction of $\vec{B}$.  The radial electric field may not lie exactly 
in this plane.  In this case the orbit plane will change when the 
electric field is applied.

Since we are interested in the spin direction relative to the momentum 
vector~\cite{bmt},  we consider the \el\ and \ma\ \f\ components 
$\vec{E}^*$ and $\vec{B}^*$ in the rest frame of the particle 
circulating in the orbit plane~\cite{bmt}.
For the closed orbit to be stable vertically, the mean vertical force in the lab frame
\begin{equation}
\langle E_{\rm v} + \beta B_r\rangle \ =\ 0.     \label{eq:EvB}  
\end{equation}

\ni Transforming to the rest frame one 
finds $\langle E_{\rm v}^*\rangle =0$, not unexpectedly because in the rest frame it 
is $\vec{E}^*$ that moves the orbit while $B^*$ generates no force. 

$B_r^*$  rotates the spin out of the orbit plane when the EDM is zero.  Using Eq.
(\ref{eq:EvB}) 

\begin{equation}
\langle B^*_r \rangle = \langle \gamma  B_r + \beta \gamma E_{\rm v} \rangle = -\langle E_{\rm v} /\beta \gamma \rangle.      \label{eq:Br}  
\end{equation}

It follows that with no electric field there is no false EDM whatever the 
shape of the orbit (error 1).
If the radial \el\ \f\ is exactly in one plane so that $\langle E_{\rm v}\rangle \ =\ 0$ 
then $\langle B_r^*\rangle $ is zero and there is again no false EDM, (error 2).

A further effect of \el\ field misalignment is that $\vec{\beta}\times\vec{E}$ 
is not parallel to $\vec{B}$  so that when (\ref{eq:C}) is satisfied there is a 
net horizontal angular velocity $\vec{\omega}_r$ acting on the spin.  However, 
if this is radially inwards on one side of the ring, it will be radially 
outwards on the other, generating a small vertical spin oscillation which 
does not accumulate from turn to turn as long as the $(g-2)$ 
precession is zero: no false EDM.  
If there is a residual $(g-2)$ precession and a radial electric field is present,
it is possible that a radial spin component can be transformed
into a vertical component.  This is the case when the radial electric field is not exactly 
orthogonal to the magnetic field.  
A single detector at a specific azimuthal location will observe a small 
EDM like signal that has the opposite sign
at a detector located $180^\circ$ apart.  The effect is proportional to the 
misalignment of the 
electric and magnetic fields from orthogonality and it goes 
to zero when the detector signals from all azimuthal locations are summed.

If the radial \el\ \f\ is not precisely in a plane (error 3) there will be 
a net vertical \el\ \f\ $\langle E_{\rm v}\rangle \ \neq\ 0$.  This will move the orbit until 
$\langle B_r^*\rangle $ satisfies Eq. \ref{eq:Br} and this will precess the spin out of 
the plane generating a false EDM.  Every precaution must be taken to 
minimize this effect but fortunately it is cancelled by alternatively injecting the 
particles clockwise (CW) and counter-clockwise (CCW) and subtracting the counts in
the detectors.  This requires 
discussion of the signs of the real and false EDM signals for $\mu^+$ 
and $\mu^-$ in each case.
The following equations indicate the signs (not magnitudes) of the real 
and false EDM angular velocities  $\vec{\omega}_e$ and  $\vec{\omega}_F$:

\begin{eqnarray}
\vec{\omega}_e & \propto &  \vec{\sigma} \times \left[ \vec{d}  \times \left( \vec{p}\times\vec{B} \right) \right] \label{eq:J} \\
\vec{\omega}_F & \propto &  \vec{\sigma} \times \left[ \vec{\mu} \times \left( \vec{p} \times \vec{E}_{\rm v} \right) \right] \label{eq:K} 
\end{eqnarray}

\ni where $\vec{\sigma}$ represents the spin vector.
 If there is a finite EDM  $\vec{d}$,  the CPT theorem requires 
$\vec{d}\cdot\vec{\sigma}$ to change sign going from $\mu^+$ to $\mu^-$.  
In Table II we show the truth table for the four different configurations 
$\mu^+$ / $\mu^-$ combined with the orbit directions CW/CCW listing the 
variables in the order 
$\vec{p},\ \vec{\sigma},\ \vec{\mu},\ \vec{d},\ \vec{B},\ \vec{E}_{\rm v}$.  We 
are displaying the situation at a fixed point in the ring, assuming that 
the muons come from pion decay in the backward direction and we arbitrarily 
make all variables positive for the reference case ($\mu^+$, CW).

\bigskip
\centerline{Table II.  Truth table for the four different configurations}
\centerline{$\mu^+/\mu^-$ combined with the orbit directions CW/CCW.}
\begin{center}
\begin{tabular}{|c|c|c|} \hline
                  & CW & CCW \\ \hline
Particle  & $\vec{p},\ \vec{\sigma},\ \vec{\mu},\ \vec{d},\ \vec{B},\ \vec{E}_{\rm v}$ & $\vec{p},\ \vec{\sigma},\ \vec{\mu},\ \vec{d},\ \vec{B},\ \vec{E}_{\rm v}$ \\ \hline
$\mu^+$   & $+,\ +,\ +,\ +,\ +,\ +$ &$ -,\ -,\ -,\ -,\ -,\ +$ \\
\hline
$\mu^-$   & $+,\ -,\ +,\ +,\ -,\ -$ &$ -,\ +,\ -,\ -,\ +,\ -$ \\
\hline
\end{tabular}
\end{center}

If the muons are emitted backward in the decay of pions in flight,  
the majority of decay electrons initially go forward (in the direction 
of $\vec{p}$), so the observed asymmetry obeys:
\begin{equation}
\vec{A}_{e,F}\ \propto \ \vec{p}\times \vec{\omega}_{e,F}\label{eq:M}.
\end{equation}

Applying equations \ref{eq:J} , \ref{eq:K} and \ref{eq:M} the results for 
real EDM asymmetry $A_e$ and false EDM asymmetry $A_F$ are listed in 
Table III.

\bigskip
\centerline{Table III.  The results for real EDM asymmetry $A_e$ and} 
\centerline{false EDM asymmetry $A_F$, applying Eqs.~(\ref{eq:J})-(\ref{eq:M}).}
\begin{center}
\begin{tabular}{|c|c|c|} \hline

          & CW & CCW \\
\hline
 Particle         & $A_e,\ A_F$ & $A_e,\ A_F$ \\
\hline
$\mu^+$   & $+,\ +$ & $-,\ +$ \\
\hline
$\mu^-$   & $+,\ +$ & $-,\ +$ \\
\hline
\end{tabular}
\end{center}

We see that the real EDM signal changes sign when the direction of 
rotation in the ring is reversed, while the false EDM due to the 
out-of-plane electric \f\ remains the same.  So this error may be 
cancelled by changing from CW to CCW if all other factors can be 
held the same.

If the \el\ \f\ is misaligned but in a plane (error 2) the orbit plane  
will change when the \el\ \f\ is applied, so the detectors, set to 
respond only to vertical spin components will include small 
contribution from horizontal spin.  This will however 
be opposite on opposite sides of the ring and so will largely cancel.
Error 4 has a similar effect.

The response of upper and lower detectors may change with time 
(error 5), so that a false asymmetry develops during each muon storage 
cycle.  Such effects can be caused by unequal detector responses to the 
changing counting rates or transients in the system triggered at 
injection time.  This effect should remain the same when muons are 
injected CW and CCW while the real EDM asymmetry changes sign.

It might be supposed that the electric and magnetic fields could be applied to 
separate sections of the orbit, with the result that the spin makes small to and 
fro movements about the vertical axis but the net $(g-2)$ precession is zero over 
one turn.  While this would fulfill the main requirement, some misalignment 
errors would not be perfectly cancelled.  For example, a harmonic of the azimuthal 
field $B_\theta $ (error 6) would cause the spin to oscillate about the horizontal 
axis parallel to $\vec{p}$.  Because rotations do not commute, the combination 
with the $(g-2)$ oscillation would generate a net rotation about the radial axis, 
leading to a false EDM.  This is an example of Berry's phase~\cite{berry84}.
Similarly, a misalignment of the electric field (error 2) would generate an 
oscillating radial angular velocity $\omega _r$ as explained above.  Combined 
with the $(g-2)$ oscillation this would give rise to a false EDM.  To minimize these 
effects, the electric and magnetic fields must be located at the same place.
This error is also canceled by injecting CW and CCW.

Further tests can be made by injecting muons with the opposite 
longitudinal polarization coming from pion decay in the forward 
direction.  In this case the maximum decay electron intensity is 
directed backwards  and all asymmetries are reversed.  But a 
false asymmetry due to detector effects (error 5) should remain the same.

Therefore, all false signals, unlike the EDM signal, will be cancelled by
 CW and CCW beam injection and by summing up the counts of all the detectors.

The method can be applied to other particles or atoms provided~\cite{khriplovich98} that 
the $g-$factor is not too far from $2$ so the 
$(g-2)$ precession can be cancelled by an accessible electric field, 
for example the deuteron.
Since the deuteron is stable, another scheme must be utilized to track the
deuteron spin.  This would most likely involve the use of an internal target in the ring.
One possible target is hydrogen gas as elastic d+p scattering is sensitive
to all of the polarization moments of the deuteron beam~\cite{glockle}. 

 In conclusion, we have
presented a new method which improves the sensitivity of EDM search for charged particles 
in storage rings by several orders of magnitude.  
It is achieved by applying an external radial electric field which  cancels
all sources of spin precession except that due to a non-zero EDM.
It is most applicable to particles with a
small anomalous magnetic moment.

This work was supported in part by the U.S. Department of Energy.  
We wish to thank the members of the EDM collaboration
for comments and criticism on the manuscript.


\begin{thebibliography}{99}
\bibitem{tinv} L.Landau,  Nucl.Phys. {\bf 3}, 127 (1957).

\bibitem{smedm} M. Pospelov and I.B. Khriplovich, Sov. J. Nucl. Phys. {\bf 53} 638 (1991).

\bibitem{pilaftsis} A. Pilaftsis, Nucl. Phys. {\bf B644},
263 (2002); 
J.L. Feng, K.T. Matchev, and Yael Shadmi, Nucl. Phys. {\bf B613},
366 (2001);
 J.R. Ellis et al., Phys. Lett. {\bf B528} 86 (2002);
 A. Romanino and A. Strumia, Nucl. Phys. {\bf B622}, 73 (2002).

\bibitem{babu}  K.S. Babu,
B. Dutta, and R.N. Mohapatra, Phys. Rev. Lett. {\bf 85}, 5064 (2000).

\bibitem{commins02} B.C. Regan {\it et al.,} 
Phys.\ Rev.\ Lett.\  {\bf 88}, 071805 (2002).


\bibitem{cern2} J. Bailey,  {\it et al.},
J. Phys. {\bf G4}, 345 (1978).

\bibitem{schiff}L.I. Schiff, Phys.Rev. {\bf 132}, 2194 (1963).

\bibitem{garled} R.L. Garwin and L. Lederman, Nuovo Cimento 11, 776 (1959).

\bibitem{Sandars_99}  
P.G.H. Sandars, Contemporary Physics 42 (2):97 (2001).

\bibitem{bmt} V. Bargmann, L. Michel, and V.L. Telegdi, Phys.\ Rev.\ Lett.\  {\bf 2}, 435 (1959).

\bibitem{jackson} J.D. Jackson, ``Classical Electrodynamics'', p. 559, $2^{\rm nd}$ ed.,
Ed. John Wiley and Sons, New York, 1975.

\bibitem{cern1} J. Bailey {\it et al.},
Nucl. Phys. B150, l (1979).


\bibitem{bnl} G.~W.~Bennett {\it et al.},  
Phys.\ Rev.\ Lett.\  {\bf 92}, 161802 (2004).

\bibitem{nelson59} D.F. Nelson {\it et al.}, Phys.\ Rev.\ Lett. {\bf 2}, 492 (1959), 
(was discovered by the authors after the current paper was accepted for publication).

\bibitem{muon_edm}M.~Aoki {\it et al}, J-PARC LOI: 
``Search for a Muon EDM at the $10^{-24} \, \rm e \cdot cm$ Level'',
$\rm http://www-ps.kek.jp/jhf-np/LOIlist/pdf/L22.pdf$
J-PARC: Japan Proton
Accelerator Research Complex, $\rm  http://j-parc.jp$


\bibitem{berry84} M.V. Berry, Proc. Roy. Soc. London {\bf A392}, 451 (1984).

\bibitem{khriplovich98} I.B. Khriplovich, Phys. Lett. {\bf B444}, 98 (1998). 

\bibitem{glockle} W. Gl\"ockle, H. Wita\l a, D. H\"uber, H. Kamada, and J. Golak,
Phys. Rep. {\bf 274}, 107 (1996).

\end{thebibliography}
\end{document}